\documentclass{article}

    \usepackage[preprint]{neurips_2018}

\usepackage[utf8]{inputenc} 
\usepackage[T1]{fontenc}    
\usepackage{hyperref}       
\usepackage{booktabs}       
\usepackage{amsfonts}       
\usepackage{nicefrac}       
\usepackage{microtype}      
\usepackage{amsmath,amsthm,amssymb,amsfonts}
\usepackage{graphicx}

\title{Fidelity Imposed Network Edit (FINE) for Solving Ill-Posed Image Reconstruction}

\author{%
  Jinwei Zhang\\
  Cornell University\\
  \texttt{jz853@cornell.edu} \\
   \And
   Zhe Liu\\
   Cornell University \\
   \texttt{zl376@cornell.edu} \\
    \And
   Shun Zhang\\
   Weill Cornell Medicine \\
   \texttt{shz2006@med.cornell.edu} \\
    \And
   Hang Zhang\\
   Cornell University \\
   \texttt{hz459@cornell.edu} \\
    \And
  Pascal Spincemaille\\
  Weill Cornell Medicine \\
  \texttt{pas2018@med.cornell.edu} \\
    \And
  Thanh D. Nguyen\\
  Weill Cornell Medicine \\
  \texttt{tdn2001@med.cornell.edu} \\
    \And
  Mert R. Sabuncu\\
  Cornell University \\
  \texttt{msabuncu@cornell.edu} \\
    \And
  Yi Wang\\
  Weill Cornell Medicine \\
  \texttt{yiwang@med.cornell.edu} \\
}

\begin{document}

\maketitle

\begin{abstract}
  Deep learning (DL) is increasingly used to solve ill-posed inverse problems in imaging, such as reconstruction from noisy and/or incomplete data, as DL offers advantages over explicit image feature extractions in defining the needed prior. However, DL typically does not incorporate the precise physics of data generation or data fidelity. Instead, DL networks are trained to output some average response to an input. Consequently, DL image reconstruction contains errors, and may perform poorly when the test data deviates significantly from the training data, such as having new pathological features. To address this lack of data fidelity problem in DL image reconstruction, a novel approach, which we call fidelity-imposed network edit (FINE), is proposed. In FINE, a pre-trained prior network’s weights are modified according to the physical model, on a test case. Our experiments demonstrate that FINE can achieve superior performance in two important inverse problems in neuroimaging: quantitative susceptibility mapping (QSM) and under-sampled reconstruction in MRI.
\end{abstract}

\section{Introduction}
Image reconstruction from noisy and/or incomplete data is often solved with regularization of various forms, which can be formulated in Bayesian inference as a maximum a posteriori (MAP) estimation (Gindi et al., 1993; Herman et al., 1979). Traditionally, these regularizations promote desired properties of explicitly extracted image features, such as image gradients or wavelet coefficients (Block et al., 2007; Fessler, 2010; Lustig et al., 2007; Uecker et al., 2008). Deep learning (DL) using a convolutional neural network (CNN) of many layers has demonstrated superior capability in capturing all desired image features than an explicit feature extraction and achieved tremendous success in a wide range of computer vision applications (Gatys et al., 2015; Johnson et al., 2016; LeCun et al., 2015; Simonyan et al., 2013). 

For DL-based image reconstruction problem, the most popular approach is to formulate it as supervised learning, where a DL model is first trained on pairs of data and its ground truth image and is then used to directly reconstruct an image from a test data. However, the performance of this supervised DL strongly depends on the similarity between test data and training data (Bickel, 2009). If the test case has a certain pathology or content that was not present in the training data, a DL model may not be able to capture the entire pathology or content (Knoll et al., 2019). DL networks do not implement the precise underlying physical model of the imaging system and are trained to output some average response to an input (Lehtinen et al., 2018). Consequently, they are known to introduce errors, particularly blurring, in the image generation process, and need to be corrected (Isola et al., 2017; Pathak et al., 2016).

One common approach to combine DL and the physical model of the imaging system is to use the DL model for defining an explicit regularization in the classical Bayesian MAP framework, typically via an L1 or L2 penalization (Aggarwal et al., 2019; Schlemper et al., 2018; Tezcan et al., 2017; Wang et al., 2016). However, traditional explicit regularization terms are known to offer imperfect feature descriptions and limit image quality in Bayesian reconstruction (Jin et al., 2017). 

The advantage of DL over explicit feature extractions may come from an explicit feature expression used during training that is buried deep in the many convolution layers through backpropagation (LeCun et al., 2015; Lee et al., 2009; Zeiler and Fergus, 2014). Accordingly, we propose to incorporate into the DL layers the physical model of test data or data fidelity defined by the discrepancy between the data measurement and the forward model of the targeted image. A method to achieve this approach is to edit the DL network weights via backpropagation according to the data fidelity of a given test data, and we refer to this method as fidelity imposed network edit (FINE). We report FINE results on two neuroimaging problems, quantitative susceptibility mapping (QSM) and MRI reconstruction from under-sampled k-space data. 

\begin{figure}[!t]
\center{\includegraphics[width=\textwidth]
        {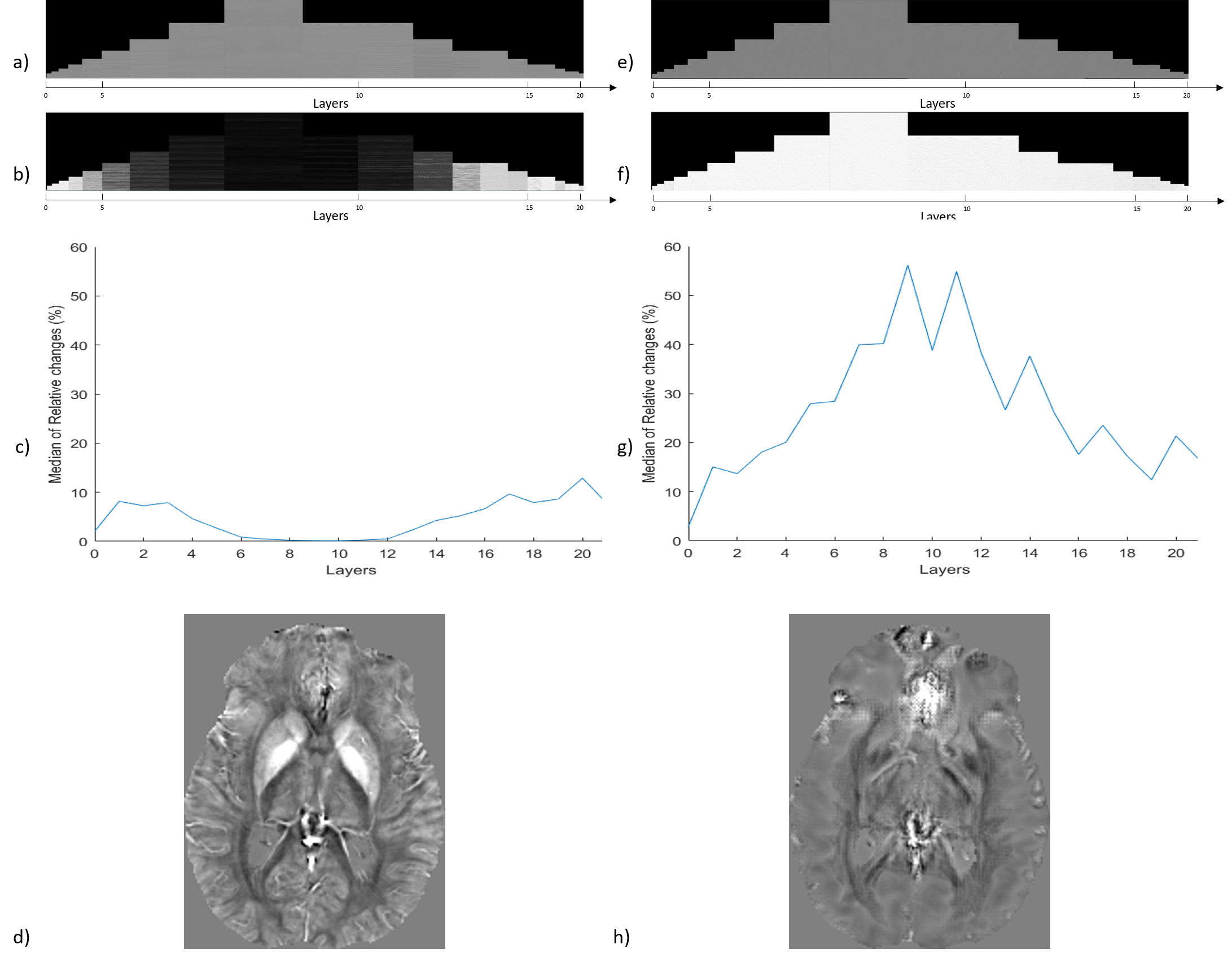}}
\caption{a) 3D U-Net’s weights for each layer before FINE. From left to right: U-Net layers from down-sampling to up-sampling sequentially with each square representing weights from a layer. b) Weight change in absolute value after FINE. The high-level layers (layers 1 through 3 and layers 16 through 20) of U-Net experienced substantial weight change by FINE. c) Median relative change of the weights in each layer after FINE. d) Reconstructed QSM after FINE. e) randomize initialization as in deep image prior, and f) corresponding weight change, showing all layers of the U-Net experienced substantial weight change. g) Median relative change of the weights in each layer after the deep prior update. h) reconstructed QSM with deep image prior failed to invert the field to susceptibility source map.}
\end{figure}

\section{Theory}
A major challenge in medical image reconstruction is to invert an ill-posed system matrix $A$ of a known physical process in the presence of data noise $n$: 
\begin{equation}
y=A x+n
\end{equation}
where $x$ is the desired image and $y$ the measured data. For example, the dipole kernel is zero on the cone surface of the magic angles, making the inversion from measured magnetic field to susceptibility source (quantitative susceptibility mapping, QSM) ill-posed; the under-sampling mask contains many zeroes, making reconstruction of under-sampled data ill-posed. Additional prior knowledge is required to obtain a solution. The Bayesian inference approach provides an optimal estimation according to measured data noise property and prior knowledge. Gaussian noise is observed in MRI complex data and may be an approximate model for various data. This leads to the common Bayesian reconstruction under Gaussian noise: 
\begin{equation}
\hat{x}=\underset{x}{\operatorname{argmin}} \frac{1}{2}\|W(A x-y)\|_{2}^{2}+R(x)
\end{equation}
where $W$ is the square root of the inverse of the noise covariance matrix, $R(x)$ is a regularization term that characterizes prior knowledge. The L2 term in Eq.2 is referred to as data fidelity. Eq.2 can be solved using numerical optimization procedures, such as the quasi-Newton method that iteratively linearizes the problem with each linear problem solved by the conjugate gradient method. Common choices for $R(x)$ include sparsity enforcement expressed as Total Variation (TV) (Osher et al., 2005) or the L1 norm in an appropriate wavelet domain (Donoho, 1995). These types of priors are critical for solving the ill-posed inverse problem. However, they can also limit image quality of the reconstruction, such as introducing artificial blockiness. 

Fundamentally, regularization promotes desired image features, which may be advantageously performed with deep learning (DL) than with conventional explicit feature extraction. A general data-to-image CNN $\phi\left(\cdot ;\Theta_{0}\right)$ with the network’s weights denoted as $\Theta_{0}$ can be trained in a supervised fashion based on $\left\{\alpha_{i}, \beta_{i}\right\}$ pairs, with $\alpha_{i}$ the ground-truth image for training data $\beta_{i}$. The weights at each convolutional layer, along with non-linear activation functions, may be regarded as a collection of feature extractors for the desired image reconstruction (Krizhevsky et al., 2012; LeCun et al., 2015). The huge number of weights in DL may explain DL’s advantage of explicit feature extraction that uses a single or few weights (Lee et al., 2009; Zeiler and Fergus, 2014). Though the training data $\beta$ in general may be different from the test data $y$ in type (sizes, contrasts, etc), one may treat a test data $y$ as the same type as the training data $\beta$ to generate a DL reconstruction:

\begin{equation}
\hat{x}=\phi\left(y ; \Theta_{0}\right)
\end{equation}

An example of such an approach applied to solve ill-posed inverse problems is QSMnet (Yoon et al., 2018), which aims to solve the field-to-susceptibility dipole inversion in QSM (de Rochefort et al., 2010).

These supervised DL networks can perform poorly if there is a mismatch between the test and training data (Bickel, 2009) (for example, when the test case has a certain pathology that is not present in the training dataset), because it is agnostic about the forward physical model of data generation well defined for the imaging system (Eq.1). The input data may not be fully used without applying the forward physical model. To address this fidelity-lacking problem, it has been proposed to treat the network output in Eq.3 as a regularization in Eq.2 using an L2 form cost to penalize the difference between the network output and the final optimized solution (Aggarwal et al., 2019; Schlemper et al., 2018; Tezcan et al., 2017):

\begin{equation}
\hat{x}=\underset{x}{\operatorname{argmin}} \frac{1}{2}\|W(A x-y)\|_{2}^{2}+\lambda\left\|x-\phi\left(y ; \Theta_{0}\right)\right\|_{2}^{2}
\end{equation}

We refer to this reconstruction as DL with L2 regularization (DLL2). The main drawback of this DLL2 approach is the use of the explicit L2 norm, which is known to be limiting (Jin et al., 2017) and may not be effective in reducing bias in the final solution towards the fixed network outcome that contains fidelity errors.

To take advantage of DL over explicit feature extraction, we propose to embed the data fidelity term deeply in all layers through backpropagation in DL networks. One method to implement this approach for reconstructing a desired image $x$ is to edit the weights in a pre-trained DL network under the guidance of data fidelity for a given test data $y$. The network $\phi(\cdot ; \Theta)$’s weights are initialized with $\Theta_{0}$ and are edited according to the test data’s physical fidelity of the imaging system: 

\begin{equation}
\widehat{\Theta}=\underset{\Theta}{\operatorname{argmin}} L(y ; \Theta)=\|W(A \phi(y ; \Theta)-y)\|_{2}^{2}
\end{equation}

Then the output of the updated network is the reconstruction of $x$ with both data fidelity and deep learning regularization: 

\begin{equation}
\hat{x}=\phi(y ; \widehat{\Theta})
\end{equation}

We refer to this approach as “fidelity imposed network edit (FINE)” for solving an ill-posed inverse problem using deep learning and imaging physics. 

\begin{figure}[!t]
\center{\includegraphics[width=\textwidth]
        {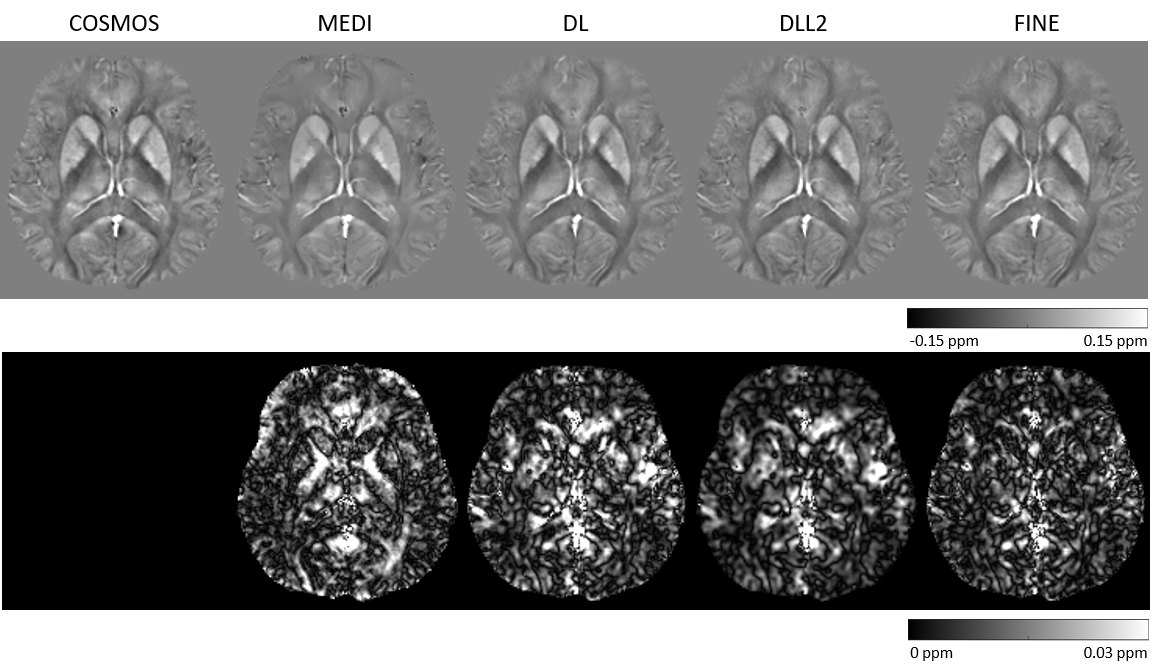}}
\caption{a) Comparison of QSM of one healthy subject reconstructed by (from left to right) COSMOS, MEDI, DL, DLL2 and FINE. First row: reconstructed QSM by four methods. Second row: absolute difference images of four methods using COSMOS as ground truth. More detailed structures were recovered after fidelity enforcement. Structures in occipital lobe were more clearly depicted in FINE and DLL2 than in MEDI and DL.}
\end{figure}

\section{Method}
In this paper, we applied the proposed FINE to two inverse problems in MRI: QSM and under-sampled k-space reconstruction. The human subject studies followed an IRB approved protocol. All images used in this work are de-identified to protect privacy of human participants. Data and code are available to all interested researchers upon request.

\subsection{QSM}
First, we applied FINE to QSM (de Rochefort et al., 2010), which is ill-posed because of zeroes at the magic angle in the forward dipole kernel. Consequently, streaking artifacts appear in the image domain after un-regularized dipole inversion (Kee et al., 2017). The Bayesian approach has been widely used to address this issue. One example is the Morphology Enabled Dipole Inversion (MEDI) method (Liu et al., 2012), which employs the following cost function:

\begin{equation}
\hat{x}=\underset{\chi}{\operatorname{argmin}} \frac{1}{2}\|W(d * \chi-f)\|_{2}^{2}+\lambda\left\|M_{G} \nabla \chi\right\|_{1}
\end{equation}

with $\chi$ the susceptibility distribution to solve, $f$ the field measurement, $d$ the dipole kernel. The regularization is a weighted total variation, with $\nabla$ the gradient operator, $M_G$ a binary edge mask determined from the magnitude image (Liu et al., 2012) which enforces morphological consistency between magnitude and susceptibility. 

\begin{figure}[!t]
\center{\includegraphics[width=\textwidth]
        {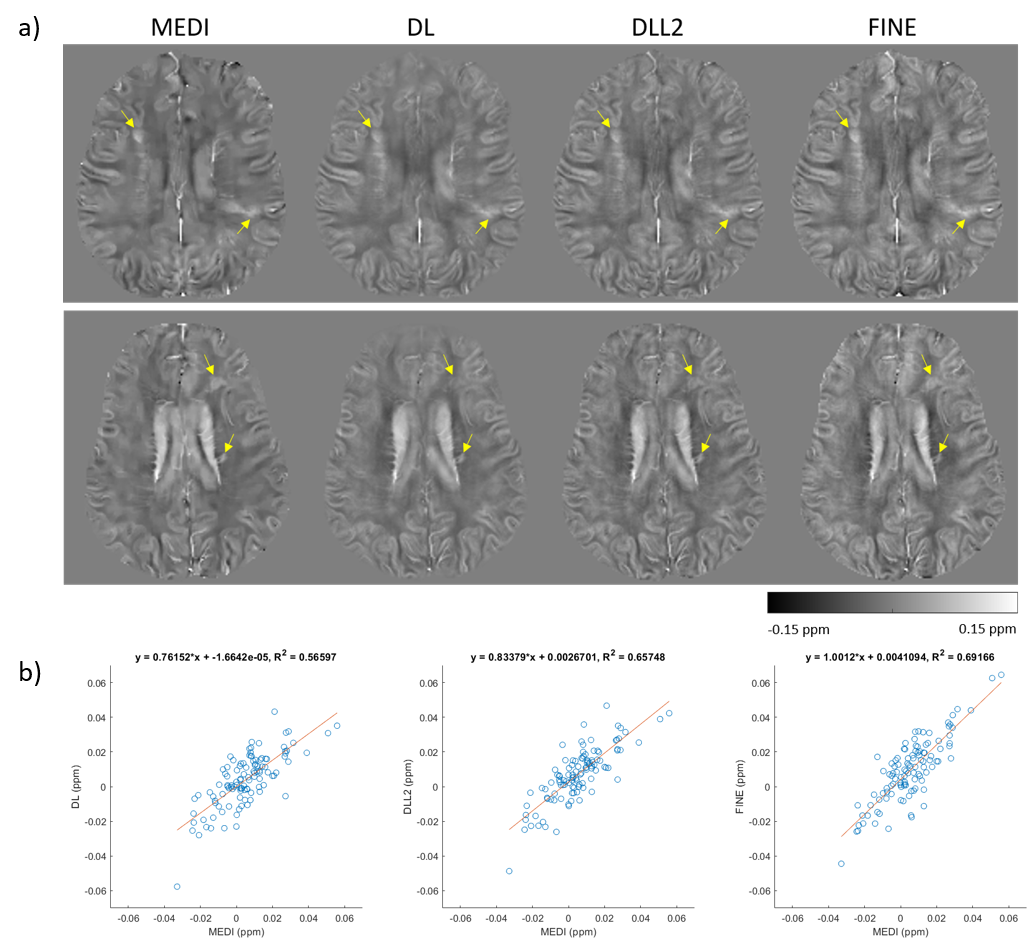}}
\caption{a) axial images from two representative MS patients. From left to right: QSMs reconstruction by MEDI, DL, DLL2 and FINE, respectively. MS lesions in QSMs were pointed out by yellow arrows. Lesions are underestimated in DL, but are recovered in DLL2 and FINE. Central veins near the ventricle were better depicted in FINE and DLL2 than in MEDI or DL. b) least square regressions of all patients’ lesion mean values between MEDI and the other three methods. FINE resolves the underestimation of lesion susceptibility seen in DL.}
\end{figure}

\subsubsection{Data acquisition and pre-processing}
MRI was performed on 6 healthy subjects using a 3T system (GE, Waukesha, WI) with a multi-echo 3D gradient echo (GRE) sequence. Detailed imaging parameters included FA = 15, FOV = 25.6 cm, TE1 = 5.0 ms, TR = 39 ms, $\#$TE = 6, $\Delta$TE = 4.6 ms, acquisition matrix = $256\times256\times48$, voxel size = $1\times1\times3$ $\text{mm}^3$, BW = $\pm \ 62.5$ kHz. The local tissue field was estimated using non-linear fitting across multi-echo phase data (Kressler et al., 2010) followed by graph-cut based phase unwrapping (Dong et al., 2015) and background field removal (Liu et al., 2011). GRE imaging was repeated at 5 different orientations per subject for COSMOS reconstruction (Liu et al., 2009), which was used as the gold standard for brain QSM. Additionally, GRE MRI was performed on 8 patients with multiple sclerosis (MS) and 8 patients with intracerebral hemorrhage (ICH) using the same 3T system with the same sequence, but only at the standard supine orientation. 

\begin{figure}[!t]
\center{\includegraphics[width=\textwidth]
        {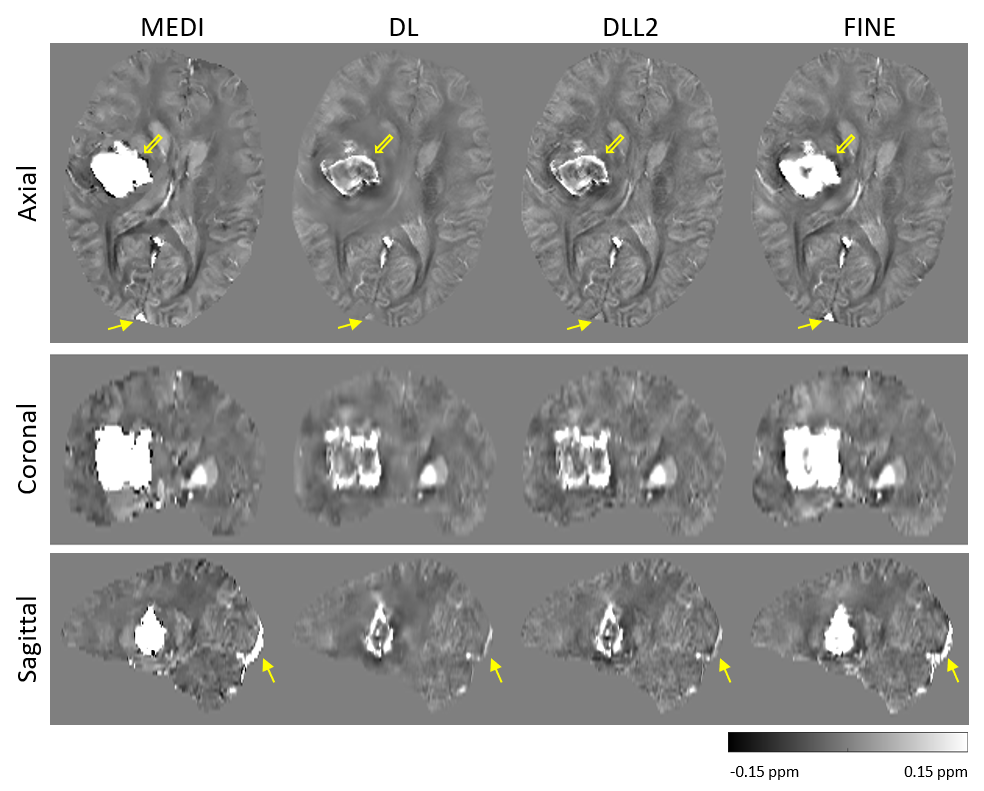}}
\caption{QSM shown in three orthogonal planes in a representative ICH patient. From left to right: QSMs reconstructed by MEDI, DL, DLL2 and FINE, respectively. Brain tissues around hemorrhage (hollow arrow) were blurry in DL, but were recovered and better depicted in DLL2 and FINE. The hemorrhagic susceptibility in DL and DLL2 was lower than in MEDI and FINE. Similarly, the susceptibility of the superior sagittal sinus (solid arrow) were underestimated in DL and DLL2, but were highlighted in MEDI and FINE.}
\end{figure}

\subsubsection{Dipole inversion network}
We implemented a 3D U-Net (Çiçek et al., 2016; Yoon et al., 2018), a fully convolutional network architecture, for mapping from the local tissue field $f$ to susceptibility distribution in QSM. The convolutional kernel size was $3\times3\times3$. 5 of the 6 healthy subjects with COSMOS QSM data were used for training, with augmentation by in-plane rotation of $\pm 15^{\circ}$. Each 3D volume data was divided into patches of size $64\times64\times32$, giving a total number of 12025 patches for training. 20\% of these patches were randomly chosen as a validation set during training. We employed the same combination of loss function as in (Yoon et al., 2018) in training the network with Adam optimizer (Kingma and Ba, 2014) (initial learning rate: $10^{-3}$, epochs: 40), resulting in a 3D U-Net $\phi\left(\cdot ; \Theta_{0}\right)$.

\subsubsection{Fidelity Imposed Network Edit (FINE)}

After pre-training the network using 3D patches described above, for a given test data, a whole local field volume $f$ was fed into the network, and the network weights $\Theta_0$ from pre-training was used to initialize the weights $\Theta$ in the following minimization:

\begin{table}[t]
  \caption{PSNR and SSIM for various QSM reconstructions in a healthy subject, with COSMOS as the ground truth reference (* denotes statistical significance for the comparison between MEDI/DL/DLL2 and FINE; p < 0.05).}
  \label{sample-table}
  \centering
  \begin{tabular}{lll}
    \toprule
         & PSNR (dB)     & SSIM \\
    \midrule
    MEDI & $44.89 \pm  \ 0.19*$  & $0.9491 \pm 
    \ 1.25\times10^{-5}*$ \\
    DL     & $45.14 \pm \ 0.40*$ & $0.9691 \pm  \ 1.36\times 10^{-5}$     \\
    DLL2  & $45.72 \pm \ 0.16$ & $0.9669 \pm \ 0.74 \times 10^{-5}*$  \\
    FINE  & $ \mathbf{46.52 \pm \ 0.50}$ & $ \mathbf{0.9720 \pm \ 0.97\times 10^{-5}}$\\
    \bottomrule
  \end{tabular}
\end{table}

\begin{table}[t]
  \caption{PSNR and SSIM for real-valued T2w MS patient test dataset reconstruction. (* denotes statistical significance for the comparison between TV/DL/DLL2/MoDL and FINE; p < 0.05).}
  \label{sample-table}
  \centering
  \begin{tabular}{lll}
    \toprule
         & PSNR (dB)     & SSIM \\
    \midrule
    TV & $38.11 \pm \ 2.62*$  & $0.9791 \pm \ 0.0090*$ \\
    DL     & $32.55 \pm \ 1.57*$ & $0.9493 \pm \ 0.0144*$     \\
    DLL2  & $37.17 \pm \ 1.78*$ & $0.9765 \pm \ 0.0078*$  \\
    MoDL & $39.42 \pm \ 1.22*$ & $0.9850 \pm \ 0.0041*$ \\
    FINE  & $ \mathbf{41.60 \pm \ 2.16}$ & $ \mathbf{0.9884 \pm \ 0.0049}$\\
    \bottomrule
  \end{tabular}
\end{table}

\begin{equation}
\widehat{\Theta}=\underset{\Theta}{\operatorname{argmin}}\|W(d * \phi(f ; \Theta)-f)\|_{2}^{2}
\end{equation}

This minimization essentially fine-tuned the pre-trained dipole inversion network to produce an output for a given test field data $f$ that would be consistent with the forward dipole model. Eq. 8 was minimized using Adam (Kingma and Ba, 2014) (initial learning rate: $10^{-4}$, iteration number: 300 (15-20 minutes per volume)). The final reconstruction of the fine-tuned network was $\hat{\chi}=\phi(f ; \widehat{\Theta})$.

FINE was applied to one healthy test subject (excluded from training), 8 MS patients and 8 ICH patients. MEDI (Liu et al., 2012) was performed with $\lambda = 0.001$ for comparison. As another benchmark, we also implemented the following based on Eq. 4:

\begin{equation}
\hat{\chi}=\underset{\chi}{\operatorname{argmin}} \frac{1}{2}\|W(d * \chi-f)\|_{2}^{2}+\lambda_{2}\left\|\chi-\phi\left(f ; \Theta_{0}\right)\right\|_{2}^{2}
\end{equation}

with $\lambda_2 = 0.01$.

\subsubsection{Quantitative analysis}
For the healthy subjects, the reconstructed QSM was compared with COSMOS (Liu et al., 2009), in terms of peak signal-to-noise ratio (PSNR) to measure the quality of reconstruction and structural similarity index (SSIM) to quantify image intensity similarity, structural similarity, and contrast similarity between pairs of image patches (Wang et al., 2004). For MS patients, least square regression of all lesion mean values across patients were employed between MEDI and the other three methods to get each pair’s linear relationship. For ICH patients, mean susceptibility values of hemorrhagic lesions on QSMs from each reconstruction method were calculated and compared. A reference-free metric to measure the blur effect of images (Crete et al., 2007) was used to quantify tissue susceptibility reconstruction quality surrounding hemorrhage (scores between 0 and 1, the less the better in terms of blur perception).

\subsection{Under-sampled reconstruction}

\begin{figure}[!t]
\center{\includegraphics[width=\textwidth]
        {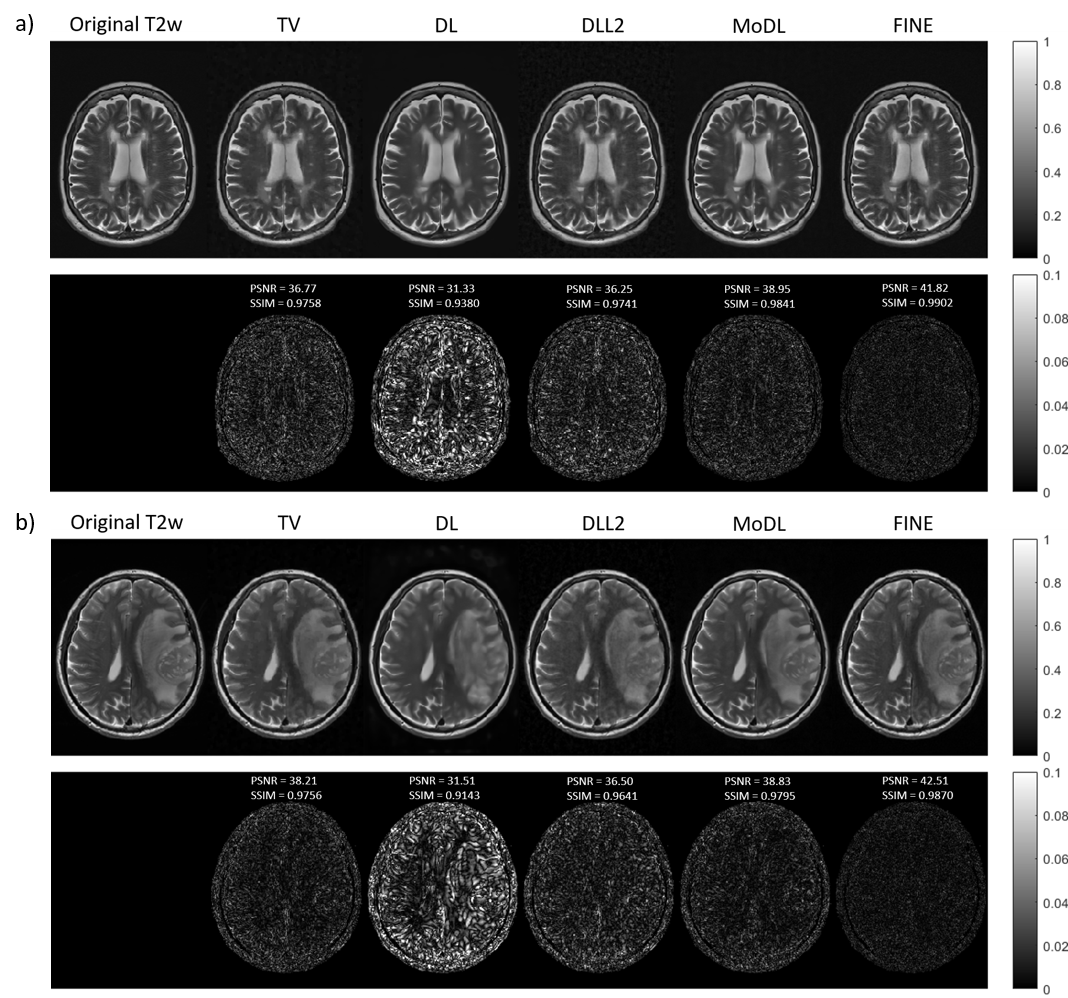}}
\caption{Reconstruction results of one MS slice in a) and one glioma slice in b). From left to right: fully sampled ground truth, under-sampled k-space reconstruction by TV, DL, DLL2, MoDL and FINE, respectively. First row: reconstructed image. Second row: magnitude of reconstruction error with respect to truth. FINE and MoDL provided a clear recovery at white/grey border and lesions, while TV suffered from over-smoothing, DLL2 suffered from noise and DL lost both structure and lesion details. FINE had the smallest reconstruction error among the five methods.}
\end{figure}

Second, we applied FINE to MRI reconstruction with under-sampled data. To accelerate the time-consuming acquisition of certain contrasts, such as T2 weighted (T2w) or T2 Fluid Attenuated Inversion Recovery (T2FLAIR) images, k-space was under-sampled, thus requiring a regularized algorithm to recover images with minimal artifact. To help with this ill-posed problem, the Bayesian approach was used for image reconstruction. Compressive Sensing MRI with Total Variation (TV) regularization is a classic approach to incorporating piece-wise constant prior of MR images to guide under-sampled reconstruction. In a single-coil Cartesian acquisition MRI in which the imaging system $A=UF$ with $U$ the binary k-space under-sampling pattern, $F$ the Fourier Transform operator, TV regularized reconstruction problem was: 

\begin{equation}
\hat{u}=\underset{u}{\operatorname{argmin}}\|U F u-b\|_{2}^{2}+\lambda\|\nabla u\|_{1}
\end{equation}

with $b$ the measured under-sampled k-space data, $b$ the image to be solved. 

\subsubsection{Data acquisition and pre-processing}
We obtained real-valued T2w axial images of 237 MS patients and 5 glioma patients, with $256\times176$ matrix size and $1 \ \text{mm}^2$ resolution. For each MS patient, we extracted 50 axial 2D image slices from each volume, giving a total number of 11850 slices. For 5 glioma patients, we extracted 44 slices with glioma tumors. Each slice was normalized to range [0, 1]. We also obtained complex-valued T2w sagittal images of 3 fully-sampled subjects used in MoDL (Aggarwal et al., 2019) as another dataset for experiments, with $256\times232$ matrix size and $1mm^2$ resolution. 

\begin{table}[t]
  \caption{PSNR and SSIM for real-valued T2w Glioma patient test set reconstruction. (* denotes statistical significance for the comparison between TV/DL/DLL2/MoDL and FINE; p < 0.05).}
  \label{sample-table}
  \centering
  \begin{tabular}{lll}
    \toprule
         & PSNR (dB)     & SSIM \\
    \midrule
    TV & $38.48 \pm \ 2.16*$  & $0.9756 \pm \ 0.0098*$ \\
    DL     & $31.79 \pm \ 1.46*$ & $0.9228 \pm \ 0.0229*$     \\
    DLL2  & $36.64 \pm \ 1.57*$ & $0.9653 \pm \ 0.0115*$  \\
    MoDL & $38.90 \pm \ 1.00*$ & $0.9798 \pm \ 0.0058*$ \\
    FINE  & $ \mathbf{41.69 \pm \ 2.00}$ & $ \mathbf{0.9845 \pm \ 0.0072}$\\
    \bottomrule
  \end{tabular}
\end{table}

\begin{table}[t]
  \caption{PSNR and SSIM for complex-valued T2w test set reconstruction. (* denotes statistical significance for the comparison between TV/DL/DLL2/MoDL and FINE; p < 0.05).}
  \label{sample-table}
  \centering
  \begin{tabular}{lll}
    \toprule
         & PSNR (dB)     & SSIM \\
    \midrule
    TV & $39.52 \pm \ 1.66*$  & $0.9867 \pm \ 0.0041*$ \\
    DL     & $28.75 \pm \ 1.95*$ & $0.9206 \pm \ 0.0259*$     \\
    DLL2  & $38.95 \pm \ 2.22*$ & $0.9853 \pm \ 0.0062*$  \\
    MoDL & $40.42 \pm \ 1.11*$ & $0.9869 \pm \ 0.0037*$ \\
    FINE  & $ \mathbf{42.85 \pm \ 2.15}$ & $ \mathbf{0.9911 \pm \ 0.0041}$\\
    \bottomrule
  \end{tabular}
\end{table}

\subsubsection{Under-sampled reconstruction network}
2D U-Net (Ronneberger et al., 2015) was used as the network architecture for mapping from $A^{H} b$ ($A^{H}(\cdot)$ maps measurement data to image domain in which U-Net works efficiently) to a fully sampled T2w image, where $U$ was chosen as a variable-density Cartesian random sampling pattern (Uecker et al., 2015). Two 2D U-Nets were employed, one for real-valued image reconstruction and the other for complex-valued image reconstruction, with complex-valued images represented by two separate real and imaginary channels in U-Net. The network was trained using $3\times3$ convolutional kernels. We used the $L_1$ difference between the network output and target image as the loss function in training the network with Adam optimizer (Kingma and Ba, 2014) (initial learning rate: $10^{-3}$, epochs: 100). For real-valued image reconstruction, 8800 slices from 176 MS patients were used for training and 2200 slices from 44 MS patients were used for validation. 850 slices from the remaining 17 MS patients and 44 slices with tumor from glioma patients formed two test datasets. Variable-density sampling pattern in real-valued dataset was generated with acceleration factor 3.24. For complex-valued image reconstruction, 2 of 3 subjects were used for training and the remaining subject was used for testing, giving 288/72/164 samples in the training/validation/test dataset, respectively. Variable-density sampling patterns used in MoDL (Aggarwal et al., 2019) with acceleration factor 6 for each coil were also applied. We used the same symbol $\phi\left(\cdot ; \Theta_{0}\right)$ to represent both trained 2D U-Nets for conciseness.

\subsubsection{Fidelity Imposed Network Edit}
Test data $b$ for a test subject was obtained by under-sampling an axial T2w image of the subject using the same sampling pattern as in the pre-training step. Similar to Eq. 8, we initialized the network weights $\Theta$ using $\Theta_0$ and updated them using the following minimization:

\begin{figure}[!t]
\center{\includegraphics[width=\textwidth]
        {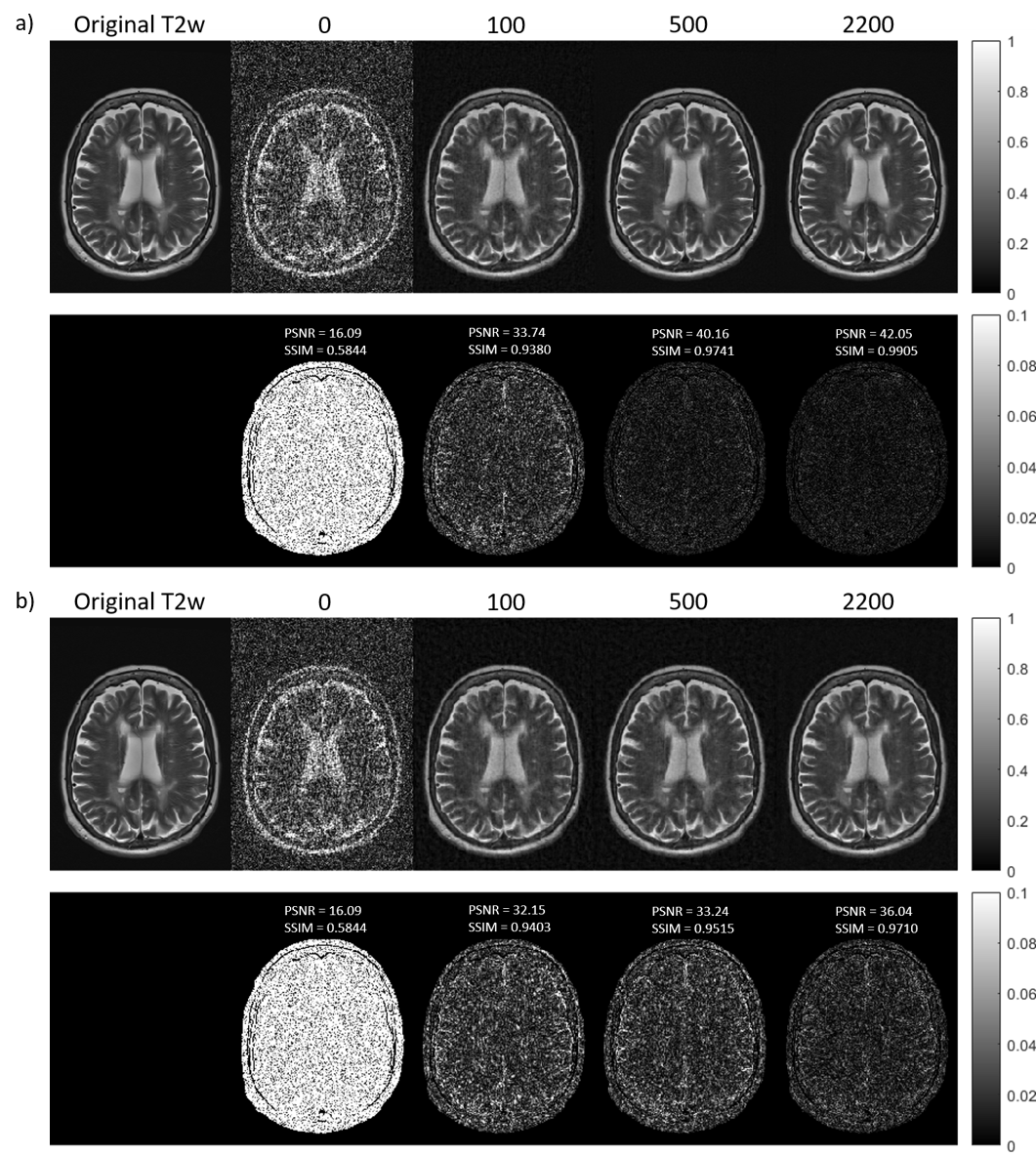}}
\caption{a) reconstruction results of FINE on one representative T2w slice with different numbers of MR images as pre-training dataset. b) reconstruction results of FINE on the same T2w slice with different numbers of natural images as pre-training dataset. Given the same size of pre-trained dataset, FINE trained on the MR image dataset had better performance than trained on the natural image dataset.}
\end{figure}

\begin{equation}
\widehat{\Theta}=\underset{\Theta}{\operatorname{argmin}}\left\|U F \phi\left(A^{H} b ; \Theta\right)-b\right\|_{2}^{2}
\end{equation}

which was solved using Adam (initial learning rate: $10^{-4}$, iteration number: 500 (2-3 minutes per slice)). FINE reconstruction for the T2w image as the final outcome of the edited network was $\hat{u}=\phi\left(A^{H} b ; \widehat{\Theta}\right)$. 

FINE reconstruction was compared with TV using $\lambda = 0.001$, and DLL2 where Eq. 4 took the following form

\begin{equation}
\hat{u}=\underset{u}{\operatorname{argmin}} \frac{1}{2}\|U F u-b\|_{2}^{2}+\lambda_{2}\left\|u-\phi\left(A^{H} b ; \Theta_{0}\right)\right\|_{2}^{2}
\end{equation}

with $\lambda_2 = 0.01$. MoDL (Aggarwal et al., 2019) was used as another benchmark for comparison, in which DLL2 in Eq. 4 was incorporated into the network structure and a series of “denoiser+DLL2” blocks were concatenated to mimic a quasi-newton optimization scheme. 

\subsubsection{Quantitative analysis}

PSNR and SSIM were calculated for TV, DLL2, MoDL and FINE to quantify the quality of reconstructed images in all experiments. To test the stability of FINE with respect to the choice of optimizer details, we repeated the experiments with different initial learning rates ($2\times10^{-4}$ and $5\times10^{-5}$) and with one additional solver (RMSprop (Tieleman and Hinton, 2012)). To test the dependency of FINE’s performance on the initial training dataset, we pre-trained multiple networks on either natural or MR images with a range of training sizes.

\section{Results}

\subsection{Edits of network weights}
The differences between $\Theta_0$ and $\Theta$ are shown in Figure 1 as an example case of applying FINE in reconstructing QSM of an MS patient. The median relative change of the weights for each layer are shown in Figures 1c\&g. FINE changed predominantly the weights in high-level layers of the U-Net (layers 1 through 3 and layers 16 through 20). Compared to FINE, a randomized $\Theta$ initialization in Eq.5, using a truncated normal distribution (Figure 1e) centered on 0 with a standard deviation = $\sqrt{2 / n}$ with $n$ the number of input units in the weight tensor (deep image prior) (Ulyanov et al., 2018), caused substantial changes of weights in all layers (Figure 1f\&g) and resulted in markedly inferior QSM (Figure 1d\&h).

\subsection{QSM}

\begin{figure}[!t]
\center{\includegraphics[width=\textwidth]
        {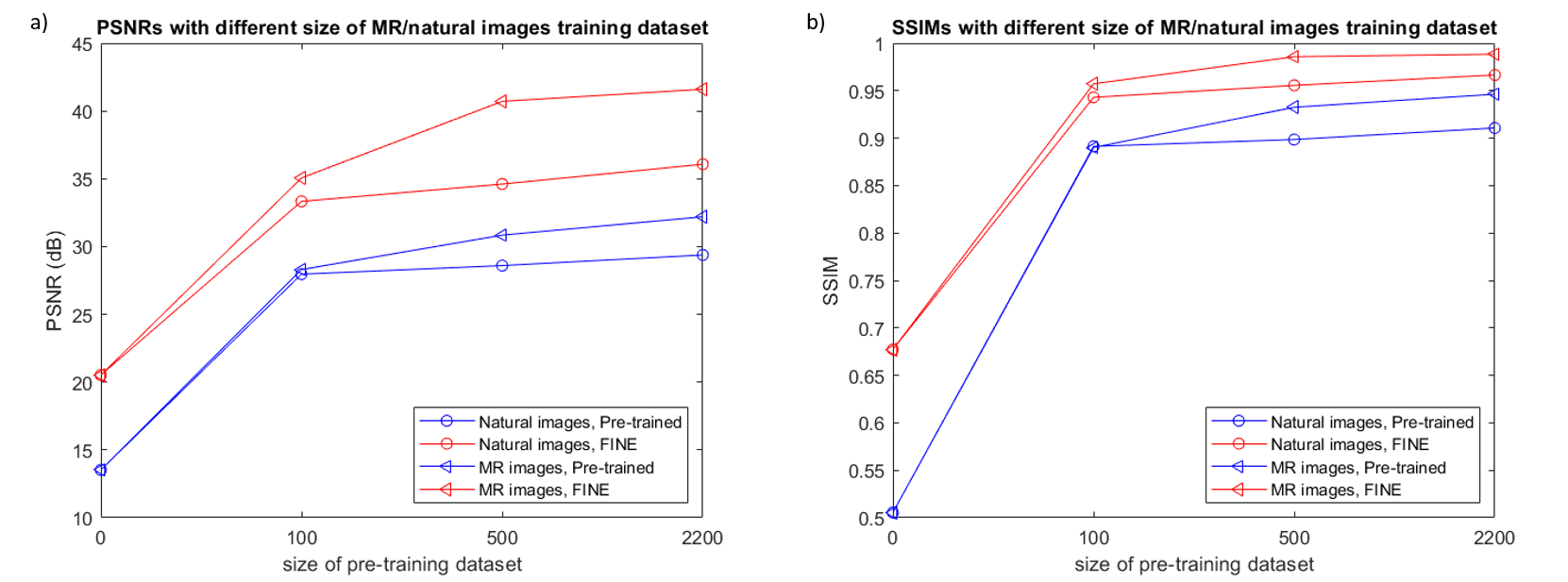}}
\caption{a) PSNR metrics of two types of pre-training dataset with different number of images before and after FINE. b) SSIM metrics of two types of pre-training dataset with different number of images before and after FINE. Given the same size of pre-trained dataset, FINE trained on MR image dataset had better performance than trained on natural image dataset. FINE with 2200 MR slices for pre-training had nearly identical performance to the one with 8800 MR slices for pre-training shown in table 2.}
\end{figure}

\subsubsection{Healthy subjects}

QSMs reconstructed by COSMOS, MEDI, DL, DLL2 and FINE are displayed in Figure 2. Structures in the occipital lobe were clearly depicted in FINE and DLL2 reconstructions, but were blurred in MEDI and DL. PSNR and SSIM in this case are shown in Table 1, with FINE demonstrating the best performance.

\subsubsection{MS patients}

QSMs reconstructed by MEDI, DL, DLL2 and FINE for two representative MS patients are displayed in Figure 3a. MS lesions were depicted using four methods (yellow arrows). Compared to MEDI, DL reconstruction underestimated the susceptibility values of certain lesions, while DLL2 and FINE managed to correct the underestimation. To quantify variations in lesion values among the different methods, mean susceptibility values of all lesions on 8 MS patients from each reconstruction method were calculated and then least square regression of all lesion mean values between MEDI and the other three methods were employed to get each pair’s linear relationship shown in Figure 3b. FINE gave better lesion estimation (slope value: 1.00) than DL (slope value: 0.76) and DLL2 (slope value: 0.83). In addition, the fine structure of central veins near the ventricle was shown more clearly on FINE and DLL2, as compared to MEDI or DL.

\subsubsection{ICH patients}
QSMs reconstructed by MEDI, DL, DLL2 and FINE for a representative ICH patient are displayed in Figure 4. Brain tissues around hemorrhages appeared blurry in the DL reconstruction but were better depicted in DLL2 and FINE. The blurring scores were $
0.18 \pm 0.02,0.23 \pm 0.03,0.18 \pm 0.02 \text { and } 0.18 \pm 0.01$ for MEDI, DL, DLL2 and FINE, respectively. MEDI, DLL2 and FINE had comparable sharpness surrounding hemorrhages, while brain tissue of DL surrounding hemorrhages were blurrier than those of the other three methods. In Figure 4, the superior sagittal sinus (solid arrow) susceptibility was lower in DL and DLL2, as compared to MEDI and FINE.

Mean susceptibility values (ppm) of hemorrhage lesions on 8 ICH patients from each reconstruction method were calculated, giving mean susceptibility values $\pm$ standard deviations: $
0.63 \pm 0.09,0.11 \pm 0.04,0.11 \pm 0.04 \text { and } 0.51 \pm 0.08 $ for MEDI, DL, DLL2 and FINE, respectively. In contrast to MEDI which showed the highest mean susceptibility values inside lesions, DL and DLL2 had considerable underestimation of lesion susceptibility, while FINE gave results close to MEDI. 

\subsection{Under-sampled reconstruction}

\subsubsection{Single-channel real-valued image reconstruction}

T2w axial images with MS lesions and glioma reconstructed by TV, DL, DLL2, MoDL and FINE are displayed in Figure 5. Structural details such as the white/grey boundary were lost in DL and were blurry in TV and DLL2. They were clearly depicted in FINE and MoDL, but with MoDL visually noisier. Lesions were also better reconstructed in FINE and MoDL. Quantitative metrics regarding PSNR and SSIM are shown in Tables 2 and 3, with FINE demonstrating the best performance.

Adam with initial learning rates $2 \times 10^{-4}$ and $5 \times 10^{-5}$, RMSprop with initial learning rate $1 \times 10^{-4}$ were also employed to test the stability of FINE on different learning rates and optimizers, resulting in average PSNRs: $41.24 \pm 2.01,41.23 \pm 1.85 \text { and } 40.82 \pm 1.60
$, and average SSIMs: $0.9876 \pm 0.0047, 0.9878 \pm 0.0047 \ \text{and} \ 0.9882 \pm 0.0040$ on the MS test dataset. As shown in Table 2, these were not significantly different from FINE either in PSNRs or SSIMs (p > 0.05).

\subsubsection{Multi-channel complex-valued image reconstruction}

Complex-valued multi-coil T2w sagittal images in test dataset were reconstructed by TV, DL, DLL2, MoDL and FINE, in which MoDL’s well-trained weights from the original paper (Aggarwal et al., 2019) were applied. Quantitative metrics regarding PSNR and SSIM are shown in Table 4, in which FINE shows the best performance.

\subsubsection{Dependency of FINE’s performance on initial training dataset}

We pre-trained several 2D U-Nets by changing the types of training dataset (single-channel real-valued MR images or natural images) and the size of the training dataset, and employed FINE using these different pre-trained weights to test how the performance of FINE depends on the initial training dataset. Figure 6 shows reconstruction results of FINE on one representative T2w slice with different numbers of MR and natural images as the pre-training dataset. Figure 7 shows the reconstruction performance on the MS test dataset in terms of PSNR and SSIM. The performance of FINE was improved as the size of the training dataset increased, either training on MR or natural images. In addition, the performance of FINE trained on natural images was below that of FINE trained on MR images, but was slightly better than end-to-end mapping trained on natural images without using FINE. It’s worth noting that FINE trained on 2200 MR images had nearly identical performance as that trained on 8800 MR images (Table 2), which indicates that FINE could reach optimal performance with less than 2200 pre-trained MR slices.

\section{Discussion}

Our results indicate that the proposed approach of fidelity-imposed network edit (FINE) can be very effective in reducing errors when applying deep learning (DL) to solving ill-posed inverse problems in medical image reconstruction. FINE embeds the desired physical model of test data in the many layers of a DL network optimized through backpropagation. Therefore, FINE realizes a physically faithful use of DL as an implicit regularization in constraining the output manifold of the DL network and offers benefits over the traditional explicit regularization in Bayesian reconstruction of ill-posed inverse problems. Compared with traditional total variation (TV) regularization, DL and a DL based L2 regularization, FINE demonstrates advantages in recovering subtle anatomy missed in TV and in resolving pathology unencountered in the DL training data.

DL has recently been used to solve inverse problems in medical image reconstruction, often using a CNN to directly map from the data domain to the image domain. For example, DL can be used to map the tissue field into QSM (Rasmussen et al., 2018; Yoon et al., 2018). This approach bypasses the typical time-consuming iterative reconstruction in traditional numerical optimization and significantly reduces the reconstruction time (down to a forward pass through the network). However, the fidelity between the reconstructed image and the actual measured data is not considered in this DL approach. This problem of lacking data fidelity has been recently recognized in CNN image reconstruction. Data fidelity may be approximately encouraged through iterative projections using many convolutional networks (Mardani et al., 2019). A more precise enforcement of data fidelity is to use the Bayesian framework with an explicit regularization, typically the L2 norm of the difference between the desired image and the network output (DLL2) (Aggarwal et al., 2019; Schlemper et al., 2018; Tezcan et al., 2017; Wang et al., 2016). However, an explicit regularization using the L2 norm or other forms can introduce artifacts in the final reconstructed image and is regarded to be inferior to DL for image feature characterization (Jin et al., 2017). In the case of measured data containing a pathological feature markedly deviating from the training data, the bias in the network output might not be effectively compensated. This is exemplified in Figure 4, where the hemorrhage feature not encountered in training datasets of healthy subjects cannot be properly captured by DL and DLL2. This problem is addressed in the proposed FINE method, where the pre-trained network bias is effectively reduced by updating the network weights guided by the measured data. Compared to DLL2 where only image voxel intensity values are optimized, many more network weights are updated in FINE, which may afford the proposed method more flexibility and greater effectiveness than DLL2. 

There are substantial neuroimaging interests in QSM (Wang et al., 2017), including studies of the metabolic rate of oxygen consumption (Zhang et al., 2015), brain tumor (Deistung et al., 2013), deep brain stimulation (Dimov et al., 2019), multiple sclerosis (Chen et al., 2014), cerebral cavernous malformation (Tan et al., 2014), Alzheimer’s disease (Acosta-Cabronero et al., 2013), Parkinson’s disease (Murakami et al., 2015), and Huntington’s disease (van Bergen et al., 2016). As QSM needs prior information to execute the ill-posed dipole inversion, seeking a better image feature for regularizing reconstruction has continuously been a major development effort (Kee et al., 2017; Langkammer et al., 2018; Wang and Liu, 2015). Mathematically, regularization should project out or suppress the streaking artifacts associated with granular noise and shadow artifacts associated with smooth model errors (Kee et al., 2017). Streaking artifacts have been effectively reduced using L1-type regularizations, but these techniques suffer from staircase artifacts or blockiness. Shadow artifacts have yet to be effectively suppressed, partly due to white matter magnetic anisotropy (Liu et al., 2018; Wisnieff et al., 2013). These QSM reconstruction challenges may be addressed more effectively using sophisticated and complex image features (Langkammer et al., 2018). DL promises to provide the desired but indescribable complex image features. The FINE implementation of DL reported here, particularly exemplified by the results in Figure 2, may realize the potential of DL for QSM reconstruction. 

Related prior work is deep image prior that trains a DL network from scratch on a single dataset for inverse problems of denoising, super-resolution, and inpainting (Ulyanov et al., 2018). Our work in Figure 1 showed that for QSM reconstruction, deep image prior fails to produce satisfying results and the use of pre-trained weights or FINE is necessary. In FINE, the network is initialized to a pre-trained network, rather than trained from scratch. Our empirical analysis indicates that FINE changes predominately the weights of initial and final (high-level) layers of U-Net for the case of QSM reconstruction in an MS patient (Figure 1), which reflects image contents specific to the patient. The effectiveness of FINE is exemplified in substantial lesions such as hemorrhages, an important QSM application to date cerebral cavernous malformation (Tan et al., 2014). Standard DL fails to reconstruct the large susceptibility values in hemorrhagic lesions and brain tissue surrounding hemorrhages, which is rectified to a large extend by FINE. However, strong susceptibility sources such as hemorrhage remain challenging due to poor signal, and the residual discrepancy between MEDI and FINE requires further investigation. 

Similar to QSM, under-sampled k-space reconstruction also requires suitable regularizations to suppress aliasing artifacts associated with the under-sampling pattern. L1-type regularizations for Bayesian inference based image reconstruction have been shown to be effective in suppressing such aliasing artifacts, but image quality suffers from blockiness. Using a neural network output as regularization (DLL2) has shown improvement in Bayesian reconstruction of data with high under-sampling rates (Aggarwal et al., 2019; Schlemper et al., 2018; Tezcan et al., 2017), and FINE promises further improvement in image reconstruction of under-sampled data, as shown in Figure 5. The under-sampling rate may be further increased as in multi-contrast MRI (Huang et al., 2012). Various magnetization preparations may be used to acquire data (Nguyen et al., 2008), and motion during data acquisition may be tracked using various navigator signals (Wang et al., 1996). As the physical model of MRI data generation is known, fast MRI using highly under-sampled rate, multiple contrasts, neural network reconstruction with FINE seems very promising in future clinical practice. 

Future work might involve assessing FINE in a wide range of applications including super-resolution and de-noising. The computational cost of FINE is much higher than a single pass through a DL network, due to the additional network updating based on the iterative optimization. The computational cost may be reduced by updating a subset of layers instead of the entire network in the optimization, as in transfer learning (Shin et al., 2016). 

In summary, data fidelity can be used to update a deep learning network on a single test dataset in order to produce high quality image reconstructions. This fidelity imposed network edit (FINE) strategy promises to be useful for solving ill-posed inverse problems in medical imaging.

\section*{Acknowledgments}
This research was supported by National Institute of Health (R01 NS090464 and S10 OD021782).

\section*{References}
Acosta-Cabronero, J., Williams, G.B., Cardenas-Blanco, A., Arnold, R.J., Lupson, V., Nestor, P.J., 2013. In vivo quantitative susceptibility mapping (QSM) in Alzheimer's disease. PLoS One 8, e81093.

Aggarwal, H.K., Mani, M.P., Jacob, M., 2019. MoDL: Model-Based Deep Learning Architecture for Inverse Problems. IEEE transactions on medical imaging 38, 394-405.

Bickel, S., 2009. Learning under differing training and test distributions.

Block, K.T., Uecker, M., Frahm, J., 2007. Undersampled radial MRI with multiple coils. Iterative image reconstruction using a total variation constraint. Magnetic resonance in medicine 57, 1086-1098.

Chen, W., Gauthier, S.A., Gupta, A., Comunale, J., Liu, T., Wang, S., Pei, M., Pitt, D., Wang, Y., 2014. Quantitative susceptibility mapping of multiple sclerosis lesions at various ages. Radiology 271, 183-192.

Çiçek, Ö., Abdulkadir, A., Lienkamp, S.S., Brox, T., Ronneberger, O., 2016. 3D U-Net: learning dense volumetric segmentation from sparse annotation. International Conference on Medical Image Computing and Computer-Assisted Intervention. Springer, pp. 424-432.

Crete, F., Dolmiere, T., Ladret, P., Nicolas, M., 2007. The blur effect: perception and estimation with a new no-reference perceptual blur metric. Human vision and electronic imaging XII. International Society for Optics and Photonics, p. 64920I.

de Rochefort, L., Liu, T., Kressler, B., Liu, J., Spincemaille, P., Lebon, V., Wu, J., Wang, Y., 2010. Quantitative susceptibility map reconstruction from MR phase data using bayesian regularization: validation and application to brain imaging. Magnetic resonance in medicine 63, 194-206.

Deistung, A., Schweser, F., Wiestler, B., Abello, M., Roethke, M., Sahm, F., Wick, W., Nagel, A.M., Heiland, S., Schlemmer, H.-P., Bendszus, M., Reichenbach, J.R., Radbruch, A., 2013. Quantitative Susceptibility Mapping Differentiates between Blood Depositions and Calcifications in Patients with Glioblastoma. PLoS ONE 8, e57924.

Dimov, A., Patel, W., Yao, Y., Wang, Y., O'Halloran, R., Kopell, B.H., 2019. Iron concentration linked to structural connectivity in the subthalamic nucleus: implications for deep brain stimulation. J Neurosurg, 1-8.

Dong, J., Liu, T., Chen, F., Zhou, D., Dimov, A., Raj, A., Cheng, Q., Spincemaille, P., Wang, Y., 2015. Simultaneous phase unwrapping and removal of chemical shift (SPURS) using graph cuts: application in quantitative susceptibility mapping. IEEE transactions on medical imaging 34, 531-540.

Donoho, D.L., 1995. Nonlinear solution of linear inverse problems by wavelet–vaguelette decomposition. Applied and computational harmonic analysis 2, 101-126.

Fessler, J.A., 2010. Model-based image reconstruction for MRI. IEEE Signal Processing Magazine 27, 81-89.

Gatys, L., Ecker, A.S., Bethge, M., 2015. Texture synthesis using convolutional neural networks. Advances in Neural Information Processing Systems, pp. 262-270.

Gindi, G., Lee, M., Rangarajan, A., Zubal, I.G., 1993. Bayesian reconstruction of functional images using anatomical information as priors. IEEE transactions on medical imaging 12, 670-680.

Herman, G.T., Hurwitz, H., Lent, A., Lung, H.-P., 1979. On the Bayesian approach to image reconstruction. Information and Control 42, 60-71.

Huang, J., Chen, C., Axel, L., 2012. Fast multi-contrast MRI reconstruction. Med Image Comput Comput Assist Interv 15, 281-288.

Isola, P., Zhu, J.-Y., Zhou, T., Efros, A.A., 2017. Image-to-image translation with conditional adversarial networks. Proceedings of the IEEE conference on computer vision and pattern recognition, pp. 1125-1134.

Jin, K.H., McCann, M.T., Froustey, E., Unser, M., 2017. Deep convolutional neural network for inverse problems in imaging. IEEE Transactions on Image Processing 26, 4509-4522.

Johnson, J., Alahi, A., Fei-Fei, L., 2016. Perceptual losses for real-time style transfer and super-resolution. European Conference on Computer Vision. Springer, pp. 694-711.

Kee, Y., Liu, Z., Zhou, L., Dimov, A., Cho, J., De Rochefort, L., Seo, J.K., Wang, Y., 2017. Quantitative susceptibility mapping (QSM) algorithms: mathematical rationale and computational implementations. IEEE Transactions on Biomedical Engineering 64, 2531-2545.

Kingma, D.P., Ba, J., 2014. Adam: A method for stochastic optimization. arXiv preprint arXiv:1412.6980.

Knoll, F., Hammernik, K., Kobler, E., Pock, T., Recht, M.P., Sodickson, D.K., 2019. Assessment of the generalization of learned image reconstruction and the potential for transfer learning. Magnetic resonance in medicine 81, 116-128.

Kressler, B., De Rochefort, L., Liu, T., Spincemaille, P., Jiang, Q., Wang, Y., 2010. Nonlinear regularization for per voxel estimation of magnetic susceptibility distributions from MRI field maps. IEEE transactions on medical imaging 29, 273-281.

Krizhevsky, A., Sutskever, I., Hinton, G.E., 2012. Imagenet classification with deep convolutional neural networks. Advances in neural information processing systems, pp. 1097-1105.

Langkammer, C., Schweser, F., Shmueli, K., Kames, C., Li, X., Guo, L., Milovic, C., Kim, J., Wei, H., Bredies, K., 2018. Quantitative susceptibility mapping: report from the 2016 reconstruction challenge. Magnetic resonance in medicine 79, 1661-1673.

LeCun, Y., Bengio, Y., Hinton, G., 2015. Deep learning. nature 521, 436.

Lee, H., Grosse, R., Ranganath, R., Ng, A.Y., 2009. Convolutional deep belief networks for scalable unsupervised learning of hierarchical representations. Proceedings of the 26th annual international conference on machine learning. ACM, pp. 609-616.

Lehtinen, J., Munkberg, J., Hasselgren, J., Laine, S., Karras, T., Aittala, M., Aila, T., 2018. Noise2noise: Learning image restoration without clean data. arXiv preprint arXiv:1803.04189.

Liu, J., Liu, T., de Rochefort, L., Ledoux, J., Khalidov, I., Chen, W., Tsiouris, A.J., Wisnieff, C., Spincemaille, P., Prince, M.R., 2012. Morphology enabled dipole inversion for quantitative susceptibility mapping using structural consistency between the magnitude image and the susceptibility map. NeuroImage 59, 2560-2568.

Liu, T., Khalidov, I., de Rochefort, L., Spincemaille, P., Liu, J., Tsiouris, A.J., Wang, Y., 2011. A novel background field removal method for MRI using projection onto dipole fields. NMR in Biomedicine 24, 1129-1136.

Liu, T., Spincemaille, P., De Rochefort, L., Kressler, B., Wang, Y., 2009. Calculation of susceptibility through multiple orientation sampling (COSMOS): a method for conditioning the inverse problem from measured magnetic field map to susceptibility source image in MRI. Magnetic Resonance in Medicine: An Official Journal of the International Society for Magnetic Resonance in Medicine 61, 196-204.

Liu, Z., Spincemaille, P., Yao, Y., Zhang, Y., Wang, Y., 2018. MEDI+0: Morphology enabled dipole inversion with automatic uniform cerebrospinal fluid zero reference for quantitative susceptibility mapping. Magn Reson Med 79, 2795-2803.

Lustig, M., Donoho, D., Pauly, J.M., 2007. Sparse MRI: The application of compressed sensing for rapid MR imaging. Magnetic resonance in medicine 58, 1182-1195.

Mardani, M., Gong, E., Cheng, J.Y., Vasanawala, S.S., Zaharchuk, G., Xing, L., Pauly, J.M., 2019. Deep Generative Adversarial Neural Networks for Compressive Sensing MRI. IEEE transactions on medical imaging 38, 167-179.

Murakami, Y., Kakeda, S., Watanabe, K., Ueda, I., Ogasawara, A., Moriya, J., Ide, S., Futatsuya, K., Sato, T., Okada, K., Uozumi, T., Tsuji, S., Liu, T., Wang, Y., Korogi, Y., 2015. Usefulness of quantitative susceptibility mapping for the diagnosis of Parkinson disease. AJNR Am J Neuroradiol 36, 1102-1108.

Nguyen, T.D., de Rochefort, L., Spincemaille, P., Cham, M.D., Weinsaft, J.W., Prince, M.R., Wang, Y., 2008. Effective motion-sensitizing magnetization preparation for black blood magnetic resonance imaging of the heart. J Magn Reson Imaging 28, 1092-1100.

Osher, S., Burger, M., Goldfarb, D., Xu, J., Yin, W., 2005. An iterative regularization method for total variation-based image restoration. Multiscale Modeling \& Simulation 4, 460-489.

Pathak, D., Krahenbuhl, P., Donahue, J., Darrell, T., Efros, A.A., 2016. Context encoders: Feature learning by inpainting. Proceedings of the IEEE conference on computer vision and pattern recognition, pp. 2536-2544.

Rasmussen, K.G.B., Kristensen, M.J., Blendal, R.G., Ostergaard, L.R., Plocharski, M., O'Brien, K., Langkammer, C., Janke, A., Barth, M., Bollmann, S., 2018. DeepQSM-Using Deep Learning to Solve the Dipole Inversion for MRI Susceptibility Mapping. Biorxiv, 278036.

Ronneberger, O., Fischer, P., Brox, T., 2015. U-Net: Convolutional Networks for Biomedical Image Segmentation. Medical Image Computing and Computer-Assisted Intervention, Pt Iii 9351, 234-241.

Schlemper, J., Caballero, J., Hajnal, J.V., Price, A.N., Rueckert, D., 2018. A deep cascade of convolutional neural networks for dynamic MR image reconstruction. IEEE transactions on medical imaging 37, 491-503.

Shin, H.-C., Roth, H.R., Gao, M., Lu, L., Xu, Z., Nogues, I., Yao, J., Mollura, D., Summers, R.M., 2016. Deep convolutional neural networks for computer-aided detection: CNN architectures, dataset characteristics and transfer learning. IEEE transactions on medical imaging 35, 1285-1298.

Simonyan, K., Vedaldi, A., Zisserman, A., 2013. Deep inside convolutional networks: Visualising image classification models and saliency maps. arXiv preprint arXiv:1312.6034.

Tan, H., Liu, T., Wu, Y., Thacker, J., Shenkar, R., Mikati, A.G., Shi, C., Dykstra, C., Wang, Y., Prasad, P.V., Edelman, R.R., Awad, I.A., 2014. Evaluation of iron content in human cerebral cavernous malformation using quantitative susceptibility mapping. Invest Radiol 49, 498-504.

Tezcan, K.C., Baumgartner, C.F., Konukoglu, E., 2017. MR image reconstruction using deep density priors. arXiv, arXiv: 1711.11386.

Tieleman, T., Hinton, G., 2012. Lecture 6.5-rmsprop: Divide the gradient by a running average of its recent magnitude. COURSERA: Neural networks for machine learning 4, 26-31.

Uecker, M., Hohage, T., Block, K.T., Frahm, J., 2008. Image reconstruction by regularized nonlinear inversion—joint estimation of coil sensitivities and image content. Magnetic resonance in medicine 60, 674-682.

Uecker, M., Ong, F., Tamir, J.I., Bahri, D., Virtue, P., Cheng, J.Y., Zhang, T., Lustig, M., 2015. Berkeley advanced reconstruction toolbox. Proc. Intl. Soc. Mag. Reson. Med, p. 2486.

Ulyanov, D., Vedaldi, A., Lempitsky, V., 2018. Deep image prior. Proceedings of the IEEE Conference on Computer Vision and Pattern Recognition, pp. 9446-9454.

van Bergen, J.M., Hua, J., Unschuld, P.G., Lim, I.A., Jones, C.K., Margolis, R.L., Ross, C.A., van Zijl, P.C., Li, X., 2016. Quantitative Susceptibility Mapping Suggests Altered Brain Iron in Premanifest Huntington Disease. AJNR Am J Neuroradiol 37, 789-796.

Wang, S., Su, Z., Ying, L., Peng, X., Zhu, S., Liang, F., Feng, D., Liang, D., 2016. Accelerating magnetic resonance imaging via deep learning. Biomedical Imaging (ISBI), 2016 IEEE 13th International Symposium on. IEEE, pp. 514-517.
Wang, Y., Liu, T., 2015. Quantitative susceptibility mapping (QSM): decoding MRI data for a tissue magnetic biomarker. Magnetic resonance in medicine 73, 82-101.

Wang, Y., Rossman, P.J., Grimm, R.C., Wilman, A.H., Riederer, S.J., Ehman, R.L., 1996. 3D MR angiography of pulmonary arteries using real-time navigator gating and magnetization preparation. Magn Reson Med 36, 579-587.

Wang, Y., Spincemaille, P., Liu, Z., Dimov, A., Deh, K., Li, J., Zhang, Y., Yao, Y., Gillen, K.M., Wilman, A.H., Gupta, A., Tsiouris, A.J., Kovanlikaya, I., Chiang, G.C., Weinsaft, J.W., Tanenbaum, L., Chen, W., Zhu, W., Chang, S., Lou, M., Kopell, B.H., Kaplitt, M.G., Devos, D., Hirai, T., Huang, X., Korogi, Y., Shtilbans, A., Jahng, G.H., Pelletier, D., Gauthier, S.A., Pitt, D., Bush, A.I., Brittenham, G.M., Prince, M.R., 2017. Clinical quantitative susceptibility mapping (QSM): Biometal imaging and its emerging roles in patient care. J Magn Reson Imaging 46, 951-971.

Wang, Z., Bovik, A.C., Sheikh, H.R., Simoncelli, E.P., 2004. Image quality assessment: from error visibility to structural similarity. IEEE transactions on image processing 13, 600-612.

Wisnieff, C., Liu, T., Spincemaille, P., Wang, S., Zhou, D., Wang, Y., 2013. Magnetic susceptibility anisotropy: cylindrical symmetry from macroscopically ordered anisotropic molecules and accuracy of MRI measurements using few orientations. Neuroimage 70, 363-376.

Yoon, J., Gong, E., Chatnuntawech, I., Bilgic, B., Lee, J., Jung, W., Ko, J., Jung, H., Setsompop, K., Zaharchuk, G., 2018. Quantitative susceptibility mapping using deep neural network: QSMnet. NeuroImage 179, 199-206.

Zeiler, M.D., Fergus, R., 2014. Visualizing and understanding convolutional networks. European conference on computer vision. Springer, pp. 818-833.

Zhang, J., Liu, T., Gupta, A., Spincemaille, P., Nguyen, T.D., Wang, Y., 2015. Quantitative mapping of cerebral metabolic rate of oxygen (CMRO2) using quantitative susceptibility mapping (QSM). Magnetic Resonance in Medicine 74, 945-952.

\end{document}